\documentclass[twocolumn]{aastex61}

\newcommand\aastex{AAS\TeX}

\shorttitle{\aastex\ 3D Boltzmann-Hydro code: III. A New method for momentum feedback}
\shortauthors{Nagakura et al.}

\begin{document}
\title{Three-Dimensional Boltzmann-Hydro Code for core-collapse in massive stars

III. A New method for momentum feedback from neutrino to matter}
\correspondingauthor{Hiroki Nagakura}
\email{hirokin@astro.princeton.edu}

\author{Hiroki Nagakura}
\affiliation{Department of Astrophysical Sciences, Princeton University, Princeton, NJ 08544, USA}

\author{Kohsuke Sumiyoshi}
\affiliation{Numazu College of Technology, Ooka 3600, Numazu, Shizuoka 410-8501, Japan}

\author{Shoichi Yamada}
\affiliation{Advanced Research Institute for Science \&
Engineering, Waseda University, 3-4-1 Okubo,
Shinjuku, Tokyo 169-8555, Japan}
\affiliation{Department of Science and Engineering, Waseda University, 3-4-1 Okubo, Shinjuku, Tokyo 169-8555, Japan}

\begin{abstract}
We present a new method for neutrino-matter coupling in multi-dimensional radiation-hydrodynamic simulations of core-collapse supernova (CCSN) with the full Boltzmann neutrino transport. The development is motivated by the fact that the accurate conservation of momentum is required for reliable numerical modelings of CCSN dynamics including a recoil of proto-neutron star (PNS). The new method is built on a hybrid approach, in which we use the energy-momentum tensor of neutrinos to compute the momentum feedback from neutrino to matter in the optically thick region while we employ the collision term in the optically thin region. In this method we utilize a general relativistic description of radiation-hydrodynamics with angular moments, which allows us to evaluate the momentum feedback from neutrino to matter without inconsistency with our Boltzmann solver. We demonstrate that the new method substantially improves the accuracy of linear momentum conservation in our CCSN simulations under reasonable angular resolutions in momentum space, alleviating the difficulty in giving the diffusion limit precisely with the discrete ordinate ($S_{n}$) method. It is the first-ever demonstration that the PNS kick can be handled directly and properly in multi-dimensional radiation-hydrodynamic simulations with the full Boltzmann neutrino transport.
\end{abstract}

\keywords{supernovae: general---neutrinos---hydrodynamics}

\section{Introduction} \label{sec:intro}
First-principles numerical modeling of core-collapse supernova (CCSN) is an indispensable approach to reveal two unresolved mechanisms: explosion and neutron star (NS) kick. These two issues are intimately related with each other and should be addressed simultaneously. One of the numerical challenges in addressing the issue is a requirement of high accuracy under multi-scale, multi-dimensional (multi-D) and multi-physics circumstances, since small numerical errors could crucially affect the final outcome. In this paper we present a novel method to improve the accuracy of linear momentum conservation in Boltzmann-radiation-hydrodynamics simulations of CCSNe.

We first give a brief overview of our CCSN code and some relevant differences from others. The most important feature in our code is that multi-D Boltzmann equations for neutrino transport are solved by discretizing the full phase space (3 in space and 3 in momentum space). The basic framework of our code was developed in \citet{2012ApJS..199...17S}, which treated the non-relativistic Boltzmann equation for a given matter profile (see also \citet{2015ApJS..216....5S}). \citet{2014ApJS..214...16N} extended it to a special relativistic treatment, in which relativistic corrections were taken into account to all orders of $v/c$ in a non-conventional way. In the same paper we also coupled the Boltzmann solver with a Newtonian self-gravity hydrodynamics, which employs essentially the same numerical method as in \citet{2011ApJ...731...80N} except for special relativistic terms. \citet{2017ApJS..229...42N} further developed a moving mesh technique to track proper motions of PNS. The technique is built on a general relativistic (GR) description of radiation-hydrodynamics, in which we use a conservative form of GR Boltzmann equation in \citet{2014PhRvD..89h4073S}. It is probably a unique approach to implement the moving mesh covariantly. More recently we entered the phase of scientific runs of axisymmetric CCSN simulations on $\sim 10$ PFLOPS K-computer \citep{2018ApJ...854..136N,Harada:2018ubo}.

 On the other hand, other state-of-the-art numerical simulations have employed some approximations for neutrino transport (see e.g., \citep{2015ApJ...807L..31L,2016MNRAS.461L.112T,2016ApJS..222...20K,2016ApJ...831...98R,2016ApJ...818..123B,2016ApJ...817...72P,2017MNRAS.472..491M,2018MNRAS.481.4786J,2018ApJ...865...81O,2018arXiv180610030P,2018arXiv180910146G,2018MNRAS.477L..80K,2018MNRAS.475L..91T,2019MNRAS.482..351V} ). Most of them take advantage of (angular) moment approaches one way or another. Integrating out the angular degrees of freedom in the Boltzmann equation, one obtains an infinite hierarchy of equations for all ranks of angular moments. For practical reasons, it is truncated at a certain level: most of the schemes normally treat moments up to the 0th or 1st order. Importantly, the 2nd angular moment is related with the energy-momentum tensor of neutrinos, and its divergence describes the conservation of energy and momentum in neutrino transport (see e.g., \citet{2011PThPh.125.1255S}). It is hence straightforward for these schemes to satisfy the conservation of energy and momentum simultaneously in simulations\footnote{Strictly speaking, the statement depends on individual numerical schemes.}.

Contrary to the moment method, Boltzmann solvers do not guarantee the conservation of energy and momentum in general. The conservative form of GR Boltzmann equation implemented in our code satisfies only the conservation of number and the conservation of the energy and momentum is violated in general once the equation is finite-differenced. In fact, the difficulty of handling multiple conservation laws (number, energy and momentum) in the Boltzmann neutrino transport has been discussed over the past few decades (see e.g., \citet{1993ApJ...405..669M,2004ApJS..150..263L,2013PhRvD..88b3011C}). Unfortunately, however, no satisfactory solution has been demonstrated in realistic CCSN simulations so far.

We found in one of our recent simulations that violation of the linear momentum conservation in the Boltzmann neutrino transport could manifest itself in hydrodynamics, i.e., artificial acceleration of PNS to unphysically large velocities. In this paper we present a novel approach to reduce the error in the momentum conservation and prevent the artificial acceleration of PNS. We will not employ the moment method but cling to solving the Boltzmann equation and change the treatment of the feedback from neutrino to matter. We will also show that the previous treatment, in which the collision term is directly integrated for the feedback, generates a systematic error in the momentum exchange between neutrino and matter in the optically thick region (see in Sec.~\ref{sec:Invcause}). We investigate the cause of the error, which, it turns out, is intimately related with the problem of the Boltzmann solver in the diffusion limit. The new method alleviates the shortcoming of the Boltzmann solver and allows us to overcome the problem with computationally feasible angular resolutions in momentum space.

This paper is organized as follows. In Sec.~\ref{sec:issues} we briefly summarize the issue of the unphysical PNS acceleration that we encountered in one of our axisymmetric CCSN simulations. Then, we investigate the cause of the PNS acceleration by systematically carrying out Boltzmann simulations for a frozen fluid background in Sec.~\ref{sec:Invcause}. In Sec.~\ref{sec:NeuTransDif} we describe the connection between the artificial PNS acceleration and the problem of the Boltzmann solver in the diffusion limit, and then we present the new method in detail in Sec.~\ref{sec:newmethod}. We examine its validity using a series of axisymmetric CCSN simulations in Sec.~\ref{sec:exam}. We summarize our conclusions in Sec.~\ref{sec:conc}. Throughout this paper, Greek and Latin subscripts denote space-time and space components, respectively. We use the metric signature of $- + + +$. Unless otherwise stated, we work in units with $c=G=1$, where $c$ and $G$ are the light speed and gravitational constant, respectively.

\section{Summary of the problem} \label{sec:issues}
In one of our latest axisymmetric CCSN simulations, we found that a PNS was accelerated to velocities more than $1000 {\rm km/s}$ by the end of the simulation ($300$ms after bounce). It should be noted, however, that we did not find such an unphysically large PNS kick in other simulations and this happens only in a special situation. In Sec.~\ref{sec:Invcause}, we investigate the cause of this anomalous PNS acceleration and also clarify the reason why other models do not have the same issue. As we shall see, the large PNS acceleration turns out to be a numerical artifact due to an error in the momentum feedback from neutrino to matter in the optically thick region. Before going into details of the analysis, we summarize the essential results of the simulation, particularly focusing on the PNS kick.

In the simulation, most parts of the numerical setup are the same as those used in \citet{2018ApJ...854..136N}. We employ one of the most realistic nuclear equations-of-state \citep{2013NuPhA.902...53T,2017NuPhA.961...78T}. Neutrino-matter interactions are based on those given in \citet{1985ApJS...58..771B} except for the recent improvements \citep{2018arXiv181209811N}. Note that we also incorporated nonisoenergetic scatterings on electrons and positrons, bremsstrahlung by nucleon collisions and electron-positron pair processes together with their inverse reactions (see e.g., \citet{2005ApJ...629..922S,2012ApJS..199...17S} and references therein). We employ a 11.2 $M_{\sun}$ progenitor in \citet{2002RvMP...74.1015W}. We adopt spherical coordinates $(r, \theta)$ covering $0 \le r \le 5000{\rm km}$ and $0^{\circ} \le \theta \le 180^{\circ}$ in the meridian section and deploy $384(r) \times 128(\theta)$ grid points. Neutrino energy space is discretized non-uniformly with $20$ energy grid points. The lowest-energy cell covers $0 \le \varepsilon \le 1{\rm MeV}$ and the rest of the mesh covers $1 \le \varepsilon \le 300{\rm MeV}$ logarithmically. In our code, the polar and azimuthal angles $(\tilde{\theta}, \tilde{\phi})$ in neutrino momentum space are locally measured from the radial direction. The angular space is covered with $10(\tilde{\theta}) \times 6(\tilde{\phi})$ grid points over the entire solid angle. The grid structure is described in the appendix of \citet{2012ApJS..199...17S}. In our Boltzmann solver, we distinguish three neutrino species: electron-type neutrinos $\nu_{\rm e}$, electron-type anti-neutrinos $\bar{\nu}_{\rm e}$ and all the others collectively denoted by $\nu_x$.

\begin{figure}
\vspace{15mm}
\epsscale{1.2}
\plotone{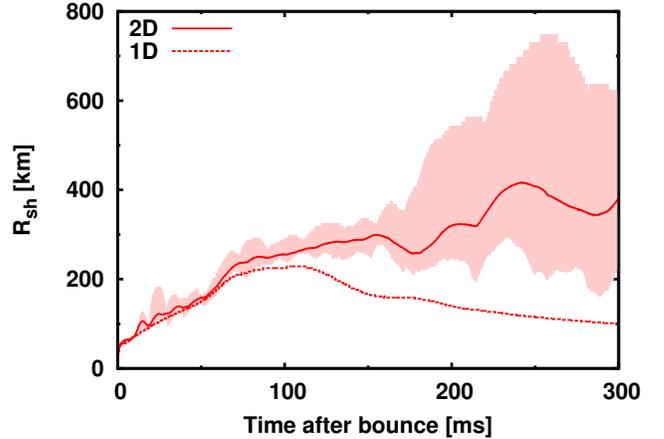}
\caption{Shock radii as a function of time in a simulation with an unphysically large accleration of PNS. The shaded region shows the range of the shock radius and the solid line represents the angle average. For comparisons, a shock radius in the spherically symmetric simulation is also displayed as dashed line.
\label{graph_rshock_Pre}}
\end{figure}

\begin{figure*}
\vspace{15mm}
\epsscale{1.2}
\plotone{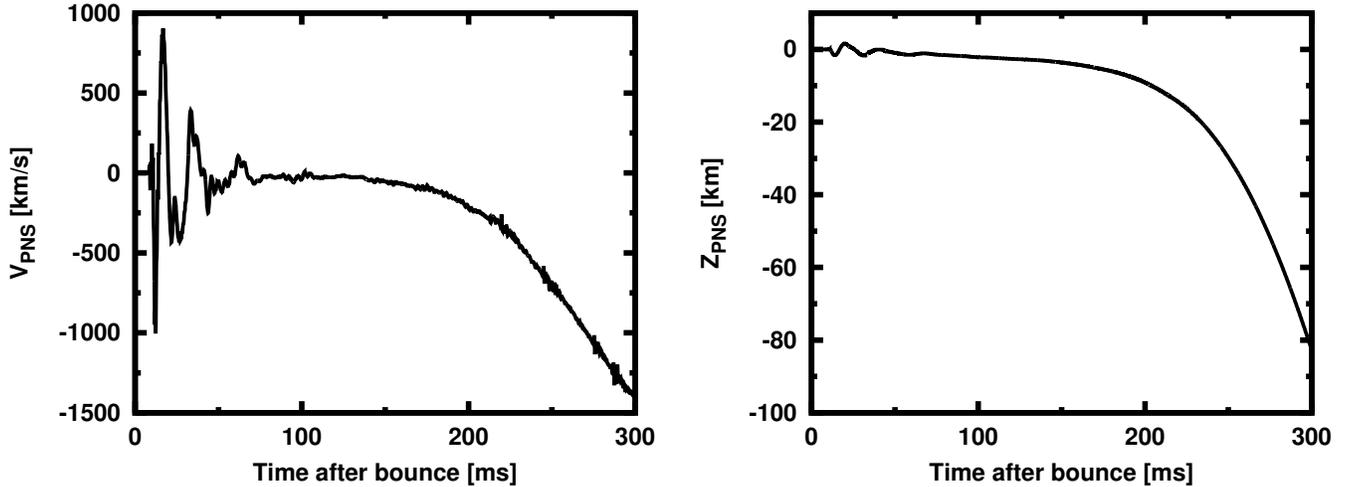}
\caption{Time evolution of the PNS velocity ($V_{\rm PNS}$) (left panel) and position along the z-axis (right panel).
\label{graph_artiPNS_velo_posi}}
\end{figure*}

In the simulation, a rapid shock expansion occurs in the northern hemisphere at $\sim 180$ms after bounce and then the north-south asymmetry persists. Figure~\ref{graph_rshock_Pre} displays the trajectory of shock radius, in which the shaded region corresponds to the range of shock radius. Then, the PNS receives a linear momentum in the opposite direction to the stronger shock expansion. We display the PNS velocity and position in Fig.~\ref{graph_artiPNS_velo_posi} as a function of time. Note that the velocity of PNS exceeds $1000 {\rm km/s}$ and still keeps rising at the end of our simulation ($t = 300 {\rm ms}$). This unphysically large acceleration is highly likely an numerical artifact.

Before we move on to the in-depth analysis of the cause of the large PNS acceleration, we would like to emphasize that the moving mesh technique developed in \citet{2017ApJS..229...42N} has nothing to do with the current issue and it works quite well in the simulation. Note that many other state-of-the-art simulations treat PNS motions quite pragmatically. For instance, some simulations sphericalize either gravitational field or matter distributions or both \citep{2012ApJ...756...84M,2014ApJ...786...83T,2015MNRAS.448.2141M,2015MNRAS.453..287M,2015ApJ...807L..31L,2016ApJ...818..123B,2017ApJ...850...43R,2018MNRAS.477.3091V,2018arXiv181105483M,2019MNRAS.482..351V}, others excise the PNS region from their computational domain \citep{2013ApJ...770...66H,2013A&A...552A.126W,2018ApJ...865...61G}.

Note that \citet{2006A&A...457..963S} implemented their own moving mesh technique to treat the proper motions of PNS. However, it was applied only to hydrodynamics but not to neutrino transport; they also excised the interior of PNS from the computational domain, in sharp contrast to our computations, in which neutrino transport and hydrodynamics are treated fully consistently on the moving mesh and the whole PNS is included in the computational domain. \citet{2010PhRvD..82j3016N} handled the NS kick directly in their radiation hydrodynamic simulations with the multi-group flux limited diffusion (MGFLD) approximation without excising the interior of PNS. However, their treatment of neutrino transport is not self-consistent: they added neutrino luminosities by hand on top of the solution of MGFLD neutrino transport in order to promote shock revival. \citet{2011ApJ...728....8B} carried out axisymmetric CCSN simulations with their own neutrino transport solver without excising the interior of PNS. They solved Boltzmann neutrino transport in the optically thin regime, meanwhile the MGFLD transport was used in the optically thick side. In addition they did not include velocity dependent terms in their transport solver. These are again sharp contrast to our computations. Note also that strong asymmetric shock expansions and NS proper motions were not observed in their simulations.

In these simulations, kick velocities of PNS have been estimated by post-process calculations, i.e., possible feedbacks from the PNS kick to neutrino transport and hydrodynamics are entirely neglected. It is true that these treatments allow CCSN simulations to avoid various practical problems arising from the PNS motion, but it is necessary to justify the prescriptions. Since our method directly handles the PNS motion in a self-consistent manner, it has a potential to serve as a reference model to validate other treatments once the current issue is addressed properly.

\section{Cause of the unphysically large acceleration of PNS} \label{sec:Invcause}

In this section, we investigate the cause of the unphysically large acceleration of PNS in detail. To identify the source of the problem clearly, we run additional simulations, in which we freeze the time evolution of matter and compute only steady neutrino distributions on top of it. As shown below, these simulations allow us to disentangle neutrino transport from the complexity of CCSN dynamics and easily quantify the error of linear momentum conservation in our Boltzmann solver. The detailed analysis finally leads us to the conclusion that the large PNS acceleration is a indeed numerical artifact caused by an error in the momentum exchange between neutrino and matter. We also find that the error is negligibly small if the matter profile inside of PNS is almost spherically symmetric. This is the main reason why other CCSN simulations did not yield such a numerical artifact. In this section, we describe these analyses in detail.

By taking the same strategy in \citet{2015ApJS..216....5S,2017ApJ...847..133R}, we compute steady neutrino distribution functions by evolving only the neutrino distributions on top of the fixed matter background in the flat spacetime and neglecting the velocity dependence. The matter profile is taken from a snapshot of our axisymmetric CCSN simulation that yielded the large PNS kick. We select the time $t=200$ms, since a runaway of PNS was initiated around this time (see Fig.~\ref{graph_artiPNS_velo_posi}). In these simulations, we also put another simplification: we do not include electron(positron)-neutrino scatterings to save the computational time. Since the opacity of these non-isoenergetic scatterings is subdominant compared with other reactions, this simplification does not affect any conclusions in this analysis.

At first, we explain a reason why the analysis on steady states is useful. We take advantage of the following fact that the advection and the neutrino-matter interactions are balanced with each other for all conserved quantities. As already explained in Sec.~\ref{sec:intro}, however, grid-based Boltzmann solvers like ours fail to guarantee all the balances simultaneously in general. In fact, since we obtain the neutrino distribution function by solving the conservative form of Boltzmann equation, the balance is satisfied for neutrino numbers but it is not the case for energy and momentum. As we will show below, we can easily quantify the violation of momentum conservation in the steady-state analysis. It also provides us with crucial information to judge whether such errors in the Boltzmann solver account for the large PNS accleration observed in the CCSN simulation.

We first prepare some useful equations to check the violation of linear momentum conservation in the steady-state. We start with the equation for the conservation law of the energy-momentum of neutrinos:
\begin{eqnarray}
&& T^{\alpha}_{({\rm rad}) \beta ; \alpha} = G_{\beta}, \label{eq:Tmunuconrad}
\end{eqnarray}
where the left- and right hand sides are defined as
\begin{eqnarray}
&& T^{\alpha}_{({\rm rad}) \beta} \equiv \sum^{}_{i=1} \int f_{\rm (i)} p^{\alpha}  p_{\beta} dV_p, \label{eq:Tmunudef} \\
&& G_{\beta} \equiv \sum^{}_{i=1} \int p_{\beta} \nu S_{\rm{rad} (\rm{i})} dV_p, \label{eq:Gdef}.
\end{eqnarray}
In these expressions, $f$ and $p^{\mu}$ denote the distribution function and four momentum of neutrinos, respectively. $S_{\rm{rad}}$ originates from the collision term for neutrino-matter interactions in the Boltzmann equations (see also Eq.~(1) in \citet{2017ApJS..229...42N}); $dV_{p} (= \nu {\rm sin}\bar{\theta} d \nu d \bar{\theta} d \bar{\phi} )$ denotes the invariant volume in the neutrino momentum space. Note that we adopt the spherical coordinates in the neutrino momentum space and $\bar{\theta}$ and $\bar{\phi}$ stand for the polar and azimuthal angles, respectively, while $\nu$ denotes the energy of neutrino. $\nu$, $\bar{\theta}$ and $\bar{\phi}$ are measured in the fluid-rest frame\footnote{We neglect the fluid-velocity dependence in all simulations discussed in this section, which implies that there is no difference between the fluid-rest and laboratory frames. Note that our new method can be applied even when the velocity-dependent terms are present. We hence specify the frame in defining the momentum variables.}. The subscript "$\rm{i}$" indicates the neutrino species.

Because of axisymmetry, we only focus on the linear momentum parallel to the symmetry axis (z-axis). We make an inner product of the unit 4-vector aligned with the z coordinate, $\mbox{\boldmath $e$}_z$, with Eq.~(\ref{eq:Tmunuconrad}):
\begin{eqnarray}
&& (\mbox{\boldmath $e$}_z)^{\beta}  T^{\alpha}_{({\rm rad}) \beta ; \alpha} = (\mbox{\boldmath $e$}_z)^{\beta} G_{\beta}. \label{eq:Tmunuconrad_Zcompo}
\end{eqnarray}
Since $\mbox{\boldmath $e$}_z$ is a killing vector in the flat spacetime, we can rewrite the equation as:
\begin{eqnarray}
&& \frac{1}{r^2 {\rm sin}{\theta}} \partial_{\alpha} 
\biggl(  
r^2 {\rm sin}{\theta}
( \mbox{\boldmath $e$}_z \cdot \mbox{\boldmath $T$} )^{\alpha}
\biggl)
= \mbox{\boldmath $e$}_z \cdot \mbox{\boldmath $G$}, \label{eq:Tmunuconrad_Zcompov2}
\end{eqnarray}
which describes the linear momentum conservation of neutrinos with respect to the z-direction. In this expression, we employ the spherical coordinates in real space since it is adopted in our Boltzmann solver. By further imposing the steady-state condition in Eq.~(\ref{eq:Tmunuconrad_Zcompov2}), it can be written as:
\begin{eqnarray}
&& \frac{1}{r^2 {\rm sin}{\theta}} \partial_{j} 
\biggl(  
r^2 {\rm sin}{\theta}
( \mbox{\boldmath $e$}_z \cdot \mbox{\boldmath $T$} )^{j}
\biggl)
= \mbox{\boldmath $e$}_z \cdot \mbox{\boldmath $G$}, \label{eq:Tmunuconrad_Zcompo_steady}
\end{eqnarray}
where $j$ stands for the spatial components: $j = \{r, \theta, \phi\}$.

\begin{figure}
\vspace{15mm}
\epsscale{1.2}
\plotone{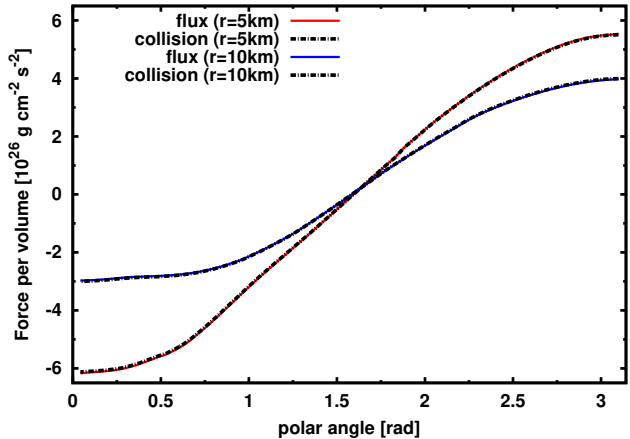}
\caption{The flux and collision terms as a function of polar angle ($\theta$) at fixed radii: $5$km (red) and $10$km (blue).
\label{graph_steadycheck_local_v2}}
\end{figure}

\begin{figure*}
\vspace{15mm}
\epsscale{1.2}
\plotone{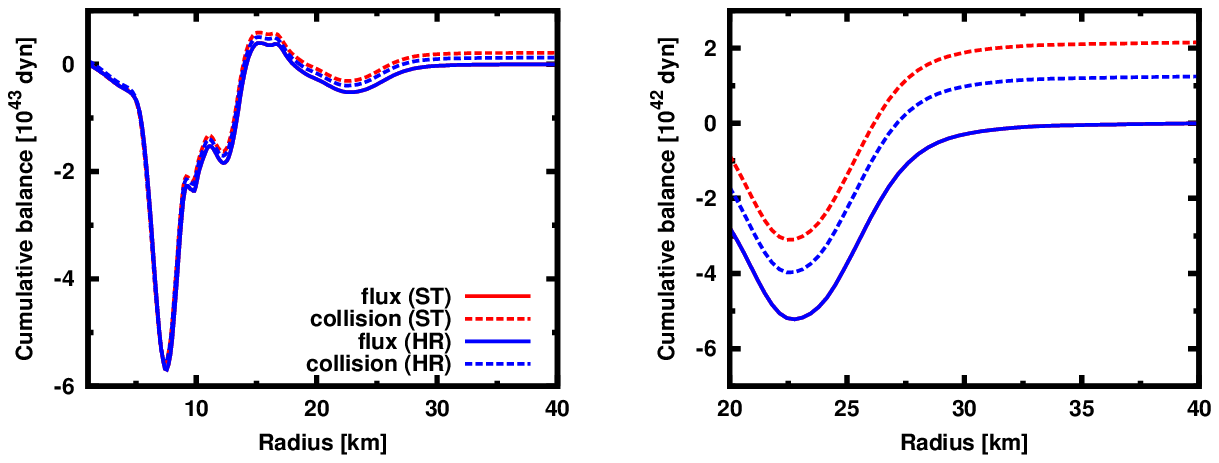}
\caption{Radial profiles of the cumulative flux- (solid lines) and collision (dashed lines) terms for the results of the steady-state Boltzmann simulations. Red and blue colors show the result in standard and high angular resolutions in momentum space (left panel) and the zoom in to the range $20 < R < 40$km (right panel). See the body of this paper for more details.
\label{graph_steady_valance}}
\end{figure*}

We evaluate each term in Eq.~(\ref{eq:Tmunuconrad_Zcompo_steady}) by using the steady profile of $f$ computed with the Boltzmann solver. We then take the sum of all terms on the left hand side of the equation, which is, hereafter, denoted by "flux-term" while the right hand side of the equation is referred to as "collision-term". Their spatial distributions are displayed in Fig.~\ref{graph_steadycheck_local_v2}. In this figure, we selected two radii inside the PNS ($r=5$ and $10$km), and show the profile as a function of the polar angle. As displayed in this figure, the two terms are almost identical at both radii, which indicates that the balance is well achieved: indeed, the difference is less than $1 \%$ with our standard angular resolution in momentum space. As we will show below, however, the precision is not sufficient to make reliable estimations of the PNS acceleration.

We proceed further to assess directly the error in the linear momentum of PNS. We take the volume integral of Eq.~(\ref{eq:Tmunuconrad_Zcompo_steady}) over the spatial range $r \le R$ to obtain the following relation:
\begin{eqnarray}
&& 2 \pi R^{2} \int_{0}^{\pi} {\rm sin}{\theta} ( \mbox{\boldmath $e$}_z \cdot \mbox{\boldmath $T$} )^{r} (R,\theta) d\theta
= \nonumber \\
&& 2 \pi \int_{0}^{R} \int_{0}^{\pi} r^2 {\rm sin}{\theta} (\mbox{\boldmath $e$}_z \cdot \mbox{\boldmath $G$}) (r,\theta) d\theta dr. \label{eq:Tmunuconrad_Zcompo_steadyInteg}
\end{eqnarray}
The left hand side of this equation represents the (radially) cumulative profile of the "flux-term" while the right hand side is the "collision-term" counterpart. The axisymmetry is also used in obtaining this expression. Importantly, the dimension in the equation is "[dyn]", which means that this equation allows us to directly quantify the error in the linear momentum of neutrinos.

In Fig.~\ref{graph_steady_valance}, the flux- (left hand side of Eq.~(\ref{eq:Tmunuconrad_Zcompo_steadyInteg})) and collision- (right hand side of Eq.~(\ref{eq:Tmunuconrad_Zcompo_steadyInteg})) terms are displayed as solid and dashed lines, respectively. Red color denotes the standard resolution. Note that we run a higher angular resolution simulation, in which we use $14(\bar{\theta}) \times 8(\bar{\phi})$ angular grid points: the result is shown in blue color. The right panel focuses on the radial range $20 \le R \le 40$km for the same quantities. Note that the radial profiles of the flux term for the two angular resolutions are almost identical: the blue solid line masks the red one. This result tells us that the flux term can be computed with a sufficiently high precision even in the standard angular resolution. On the other hand, some angular-resolution dependence can be clearly seen in the collision term. Importantly, the difference between the flux and collision terms is smaller for the higher angular resolution than for the standard one, which is the evidence that the error of linear momentum is smaller in the higher angular resolution in momentum space as expected.

The angular-resolution study reveals that the collision term contains the dominant error in the linear momentum conservation. Since the neutrino transport and hydrodynamics are coupled through the collision term in our CCSN simulations, the errors may be the cause of the PNS acceleration. Here we quantify to what extent the errors are transferred from neutrino to matter using the result of the steady-state simulations. As shown in the right panel of Fig.~\ref{graph_steady_valance}, the difference between the flux and collision terms at the edge of PNS surface ($\sim 40$km) is an order of $\sim 10^{42}$dyn, which corresponds to the PNS acceleration of $\sim 10^{9} {\rm cm/s}^2$. This value is consistent with the PNS motion we found in our CCSN simulation (see Fig.~\ref{graph_artiPNS_velo_posi}). We hence conclude that the strong PNS acceleration is a numerical artifact caused by the errors in the momentum exchange between neutrino and matter.

We performed the same analysis for the axisymmetric steady-state simulation with the spherically symmetric matter profile and confirmed that both the flux and collision terms in Eq.~(\ref{eq:Tmunuconrad_Zcompo_steadyInteg}) were zero to the machine precision. This implies that the artificial PNS acceleration appears once the matter profile strongly deviates from spherical symmetry, which occurs almost at the same time when the shock on the north side expands rapidly in the simulation that gives rise to the unphysically large acceleration of PNS . It is also important to note that the error may have increased with time due to the fact that the PNS motion itself enhances the asymmetry in the matter distribution. In other CCSN simulations, on the other hand, we found less asymmetric matter profiles in the post-shock flow including the PNS. This is the main reason why we did not encounter the numerical artifact of large PNS acceleration in these models.

The analysis in this section shows that the inaccuracy in the momentum conservation is primarily due to the lack of angular resolution in momentum space. This study also indicates that just increasing the angular resolution in momentum space will suppress the artificial acceleration of PNS (see Fig.~\ref{graph_steady_valance}). However, it is not the best solution at the moment, since the CCSN simulations with the full Boltzmann neutrino transport are numerically expensive even under the current standard resolution. In addition, we need to know how fine an angular resolution is required prior to the beginning of simulations, since it may depend on models.

\section{Neutrino transport in the diffusion regime} \label{sec:NeuTransDif}
The steady-state analysis in the previous section reveals that the large PNS acceleration is due to the violation of linear momentum conservation in the optically thick region ($r \lesssim 10$km). We also find that the flux-term defined as the left hand side of Eq.~(\ref{eq:Tmunuconrad_Zcompo_steadyInteg}) is more accurate than the collision-term for the same angular resolution. We take advantage of this property in the new method (see Sec.~\ref{sec:newmethod} for details of the method). Before entering into details of the new method, we describe how the different resolution dependences of the flux- and collision-terms are obtained in our Boltzmann solver.

It should be noted that it is not easy for the angular-grid-based Boltzmann solvers to accurately handle the neutrino transport in the diffusion regime: more specifically, it is difficult to evaluate the neutrino flux (i.e., the 1st moment) precisely (see e.g., \citet{1993ApJ...405..669M,2004ApJS..150..263L}). In our previous paper \citep{2012ApJS..199...17S}, we conducted detailed tests of our Boltzmann solver in the diffusion regime, in which we studied 1D, 2D and 3D diffusions of a Gaussian packet in a uniform matter background and found in fact that our Boltzmann solver tends to overestimate the neutrino flux due to the numerical diffusion arising from the finite resolution. As shown below, the different angular-resolution dependences of the flux- and collision-terms are intimately related with this problem of the Boltzmann neutrino transport in the diffusion regime.

In this regime, neutrinos interact with matter very frequently and, as a consequence, their angular distribution is almost isotropic. This fact allows us to express the 1st ($H_i$) and 2nd ($K_{i j}$) moments in terms of the 0th moment ($J$) as follows (see, e.g., \citet{2011PThPh.125.1255S}):
\begin{eqnarray}
&& H_{i} = \frac{1}{3 \kappa} \nabla_{i} J + \mathcal{O}({\frac{1}{\kappa^2}}), \label{eq:1stmomDif} \\
&& K_{i j} = \frac{1}{3} \gamma_{ij} J + \mathcal{O}({\frac{1}{\kappa^2}}), \label{eq:2ndmomDif}
\end{eqnarray}
where $\kappa$ and $\gamma_{i j}$ denote the total opacity and the 3-metric for flat space, respectively. From these equations, we can also obtain the following relation:
\begin{eqnarray}
\nabla_{z} K_{z z} = \kappa H_{z} + \mathcal{O}({\frac{1}{\kappa}}). \label{eq:1st2ndrelaDif}
\end{eqnarray}

It is important to realize that this is the same as the diffusion limit of Eq.~(\ref{eq:Tmunuconrad_Zcompo_steady})\footnote{See also \citet{2012arXiv1212.1623B} to derive the diffusion equation from the Boltzmann equation.}. By virtue of this correspondence, we can understand the angular-resolution dependences of the flux- and collision-terms as follows.

Eq.~(\ref{eq:1st2ndrelaDif}) indicates that the flux- and collision-terms in Eq.~(\ref{eq:Tmunuconrad_Zcompo_steadyInteg}) are nothing but the 2nd and 1st moments, respectively, in the diffusion limit. This implies that the collision-term in Eq.~(\ref{eq:Tmunuconrad_Zcompo_steadyInteg}) is more difficult to evaluate precisely with the angular-grid-based Boltzmann solvers in the diffusion limit, since we derive it mainly from the 1st moment. To the contrary, the 0th and 2nd moments are more accurate, since they are dominated by the isotropic component of the angular distribution in this regime. This is the main reason why the larger angular-resolution dependence is observed for the collision-term than for the flux-term. It should be now clear that the use of the 2nd moment instead of the 1st one for the evaluation of the momentum exchange between neutrino and matter in the diffusion regime is the key to the solution of the large-kick problem.

\section{New method} \label{sec:newmethod}

The analyses in Sections~\ref{sec:Invcause} and \ref{sec:NeuTransDif} reveal the problem in the original treatment of the momentum feedback from neutrino to matter. The direct integration of the collision term is essentially equivalent to utilizing the 1st moment of neutrino angular distribution in the diffusion regime. The angular-grid-based Boltzmann solvers like ours have difficulties in reproducing the 1st moment accurately, compared with the 0th and 2nd moments. It is important to note that the above problem have been already recognized in some previous papers: for instance, \citet{2016ApJ...817...72P} does not use the collision term but compute the neutrino-pressure gradient instead to evaluate the momentum feedback from neutrino to matter in their simulations. It should be noted that we can not utilize the same formulation to address the current issue, since their prescription is based on the diffusion approximation, which is not compatible with the Boltzmann solver. Moreover, our Boltzmann solver is formulated in the fully general relativistic framework and it is a highly non-trivial issue how to evaluate the momentum feedback without inconsistency. In this section, we present a new method, which satisfies all requirements stated above.

The essence of our new method is as follows. We modify only the treatment of the momentum feedback from neutrino to matter without changing the numerical algorithms to solve the Boltzmann equation, the hydrodynamic equations and the Poisson equation for Newtonian gravity (see Fig. 6 in \citet{2014ApJS..214...16N}). In the original treatment, we first compute the momentum exchange from individual neutrino-matter interactions, and then their sum is given to the source term of the Euler equation (see Sec.~4 in \citet{2017ApJS..229...42N}\footnote{It should be noted that there are several typos in \citet{2017ApJS..229...42N}. On the right hand side of Eq.(31), $G_{i}$ should be replaced with $\alpha \sqrt{\gamma} G_i$. In Eq.(43), $G^{r}, G^{\theta}$ and $G^{\phi}$ should be replaced with $G_{r}, G_{\theta}$ and $G_{\phi}$, respectively. Finally, the sign on the right hand side of Eq.(44) is negative.}). In the new method, on the other hand, we evaluate the momentum exchange by taking an appropriate average of the values obtained in two different ways, one of which is the original treatment and the other is the employment of the energy-momentum tensor of neutrinos (see below for more details).

 In the new method, the latter treatment is adopted only in the optically thick region with the original treatment still being used in other regions. There are mainly two reasons why we take this approach. The first one is that, in the optically thin region, the accuracy of the 0th and 2nd moments is not much different from that of the 1st moment. In fact, \citet{2017ApJ...847..133R} showed that these moments computed by our Boltzmann solver with the standard angular resolution deviate from those provided by the Monte Carlo transport (which is capable of evaluating moments with higher precision than our method) by $\sim 10 \%$ in the optically thin region. The second reason is that the original treatment is preferable for hydrodynamics in the optically thin region. This is simply because the original treatment guarantees that the feedback to hydrodynamics becomes small in the optically thin region just as expected while the new treatment may generate some artificial momentum exchange from numerical errors. Although the resolution dependence and the effect of momentum feedback on CCSN dynamics should be investigated in detail even under the original treatment to make reliable CCSN models, it is not a main subject in this study and will be conducted elsewhere.

Now we describe the new method in detail. We emphasize that the formulation is consistent with the fully general relativistic treatment of neutrino transport in our code and ensures the accurate momentum feedback from neutrino to matter even in the diffusion regime. We start with the total energy-momentum conservation:
\begin{eqnarray}
&& T^{\alpha}_{({\rm tot}) \beta ; \alpha} = 0, \label{eq:totTmunucon}
\end{eqnarray}
where the total energy-momentum tensor is decomposed as
\begin{eqnarray}
&& \mbox{\boldmath $T$}_{\rm (tot)} = \mbox{\boldmath $T$}_{\rm (rad)} + \mbox{\boldmath $T$}_{\rm (mat)}.  \label{eq:totTmunu}
\end{eqnarray}
In the above equation, $\mbox{\boldmath $T$}_{\rm (mat)}$ denotes the energy-momentum tensor of matter. From Eqs.~(\ref{eq:Tmunuconrad}),~(\ref{eq:totTmunucon}) and ~(\ref{eq:totTmunu}), we can obtain the following relation:
\begin{eqnarray}
&& T^{\alpha}_{({\rm mat}) \beta ; \alpha} = - G_{\beta}. \label{eq:GeneTmunumatcon_v1}
\end{eqnarray}
On the other hand, we can simply rewrite Eq.~(\ref{eq:totTmunucon}) as
\begin{eqnarray}
&& T^{\alpha}_{({\rm mat}) \beta ; \alpha} = - T^{\alpha}_{({\rm rad}) \beta ; \alpha} . \label{eq:GeneTmunumatcon_v2}
\end{eqnarray}
Mathematically speaking, Eqs.~(\ref{eq:GeneTmunumatcon_v1}) and~(\ref{eq:GeneTmunumatcon_v2}) are equivalent to each other. It is not true in practice, though, since Eq.~(\ref{eq:Tmunuconrad}) is not exactly satisfied in our Boltzmann solver (see in Sec.~\ref{sec:intro}). Hence, in the new method, we mix the two equations by introducing a parameter $\lambda$ $(0 \le \lambda \le 1)$,
\begin{eqnarray}
&& T^{\alpha}_{({\rm mat}) \beta ; \alpha} = - \lambda \hspace{0.5mm} T^{\alpha}_{({\rm rad}) \beta ; \alpha} - (1-\lambda) \hspace{0.5mm} G_{\beta} . \label{eq:GeneTmunumatcon_v3}
\end{eqnarray}
It is chosen so that $\lambda \rightarrow 1$ in the optically thick limit whereas $\lambda \rightarrow 0$ in the opposite limit. Since the optical depth is correlated with the matter density, we determine $\lambda$ as a function of $\rho_{\rm ave}$, which denotes the angle average of matter density. According to the steady-state analyses in Sec.~\ref{sec:Invcause}, the main region violating the balance between the flux and collision terms is the interior of PNS with $\rho_{\rm ave} \gtrsim 10^{13} {\rm g/cm}^3$. We hence take the following function:
\begin{eqnarray}
&& \lambda(\rho_{\rm ave}) = {\rm max} \left( {\rm min}(1,\frac{\rho_{\rm ave} - \rho_L}{\rho_H - \rho_L}),0 \right), \label{eq:lambdafuncrho}
\end{eqnarray}
where we set $\rho_H = 10^{13} {\rm  g/cm}^3$ and $\rho_L = 7 \times 10^{12} {\rm  g/cm}^3$, respectively. Although there may be better choices and it is possible to change $\rho_H$ and $\rho_L$ depending on the angular resolution, we do not explore them further here. The important thing is that the new method with the above choice significantly improves the accuracy of the momentum conservation and addresses well the issues of the unphysical acceleration of PNS (see Sec.~\ref{sec:exam}).

Using Eq.~(\ref{eq:GeneTmunumatcon_v3}), we rewrite the Euler equation in the new method as:
\begin{eqnarray}
&& \gamma^{\beta}_{\hspace{1.0mm}i} T^{\alpha}_{({\rm mat}) \beta ; \alpha} = 
- \lambda \hspace{0.5mm} \gamma^{\beta}_{\hspace{1.0mm}i} T^{\alpha}_{({\rm rad}) \beta ; \alpha} - (1-\lambda) \hspace{0.5mm} \gamma^{\beta}_{\hspace{1.0mm}i} G_{\beta} . \label{eq:Eulerequ_newmethod}
\end{eqnarray}
where $\gamma^{\beta}_{\hspace{1.0mm} \alpha}$ denotes the projection tensor restricted to the spacelike 3-dimensional hypersurface. Below, we describe a procedure to compute $\gamma^{\beta}_{\hspace{1.0mm}i} T^{\alpha}_{({\rm rad}) \beta ; \alpha}$ (the first term of right hand side of the equation), while we refer the reader to \citet{2017ApJS..229...42N} for the procedure computing the 2nd term. It is convenient to use a moment formalism of \citet{2011PThPh.125.1255S}, since it is fully general relativistic and is compatible with our Boltzmann solver:
\begin{eqnarray}
\gamma^{\beta}_{\hspace{1.0mm}i} T^{\alpha}_{({\rm rad}) \beta ; \alpha}  &=& \frac{1}{\alpha \sqrt{\gamma}} \biggl[ \partial_t( \sqrt{\gamma} F_i  ) + \partial_j \{  \sqrt{\gamma} ( \alpha P_{i}^{\hspace{0.5mm} j} - \beta^{j} F_{i} ) \} \biggl] \nonumber \\
 &+& \frac{1}{\alpha} ( E  \partial_i \alpha - F_{k} \partial_i \beta^{k} )
 - \frac{1}{2} P^{jk} \partial_{i} \gamma_{jk},
\label{eq:momentumexpression}
\end{eqnarray}
where $\alpha$, $\beta^{i}$ and $\gamma$ denote the lapse function, the shift vector and the determinant of 3-metric ($\gamma_{ij}$), respectively. Other quantities, $E, F_i$ and $P_{i}^{\hspace{0.5mm} j}$ are defined as
\begin{eqnarray}
&&E = T^{\alpha \beta}_{({\rm rad})} n_{\alpha} n_{\beta}, \label{eq:Edef} \\
&&F_{i} = - T^{\alpha \beta}_{({\rm rad})} n_{\alpha} \gamma_{\beta i}, \label{eq:Fidef} \\
&&P_{i}^{\hspace{0.5mm} j} = T^{\alpha \beta}_{({\rm rad})} \gamma_{\alpha i} \gamma_{\beta}^{\hspace{1.0mm} j}, \label{eq:Pijdef} 
\end{eqnarray}
where $n^{\alpha}$ denotes the unit 4-vector normal to the spacelike 3-dimensional hypersurface. We evaluate each term on the right hand side of Eq.~(\ref{eq:momentumexpression}) in our code as follows.

Suppose we know the distribution function $f$ at $t=t^{(n)}$ and $t^{(n+1)}$, where the subscripts represent the time steps. $F_{i}$ can be obtained at both time steps by using Eq.~(\ref{eq:Fidef}). We then evaluate the time derivative of $F_i$ as
\begin{eqnarray}
&& \partial_t F_{i} = \frac{ F_{i}^{(n+1)} - F_{i}^{(n)} }{\Delta t}, \label{eq:dFidt} 
\end{eqnarray}
where $F_i^{(n)}$ and $F_i^{(n+1)}$ denote the values of $F_i$ at $t=t^{(n)}$ and $t^{(n+1)}$, respectively; $\Delta t$ is the time difference between the two time steps. Note that $\partial_t \gamma$ is zero in our current CCSN simulations, since we assume that the spacetime is flat. In a dynamical spacetime, on the other hand, $\partial_t \gamma$ will be given by solving the Einstein equation possibly under the 3+1 formalism, which we will couple to our CCSN code in the future. The lapse function $\alpha$ is also set to 1 in the current CCSN code. Note that the shift vector $\beta^{i}$ is not zero and time-dependent quantity in our code since we utilize it to deal with the motion of PNS (see \citet{2017ApJS..229...42N}). Only the value of $\beta^{i}$ at $t=t^{(n)}$ is used to evaluate $\beta$-related terms on the right hand side of Eq.~(\ref{eq:momentumexpression}). The values of $E, F_i$ and $P_{i}^{\hspace{0.5mm} j}$ in the same equation are evaluated at $t=t^{(n+1)}$, on the other hand. All spatial derivatives are evaluated as the central difference.

We expect that the above formulation will improve not only the exchange of linear momentum between neutrino and matter but also the total momentum conservation in the radiation-hydrodynamic simulations. This is because our new method is essentially reduced to solving the total momentum conservation equation in the optically thick region, where $\lambda$ is set to 1 in Eq.~(\ref{eq:Eulerequ_newmethod}). In addition, the dominant component on the right hand side of Eq.~(\ref{eq:momentumexpression}) is the neutrino pressure gradient, i.e.,
\begin{eqnarray}
\gamma^{\beta}_{\hspace{1.0mm}i} T^{\alpha}_{({\rm rad}) \beta ; \alpha}  &\sim& \frac{1}{\alpha \sqrt{\gamma}} \partial_j \{  \alpha \sqrt{\gamma} P_{i}^{\hspace{0.5mm} j}  \}
 - \frac{1}{2} P^{jk} \partial_{i} \gamma_{jk},
\label{eq:momentumexpression_primary}
\end{eqnarray}
in the optically thick region, as long as the motion of PNS is slow, $V_{\rm NS}/c \ll 1$, where $V_{\rm NS}$ denotes the speed of the proper motion of PNS with respect to the laboratory frame. Recall that the neutrino pressure is evaluated much more accurately than the flux in the diffusion regime with our code. Note also that Eq.~(\ref{eq:momentumexpression_primary}) is essentially the same as the treatment of momentum feedback in \citet{2016ApJ...817...72P}. This implies that our new formulation is an extension from the previous treatment that reproduces the diffusion limit but incorporates the all other effects such as deviations from the limit as well as proper motions of PNS, which could not be handled by the previous prescription, without losing the consistency with the general relativistic Boltzmann neutrino transport.

Note also that our method is fundamentally different from the two-moment method, since we evaluate $P_{i}^{\hspace{0.5mm} j}$ based on the solution of Boltzmann equation\footnote{Note that the closure relation in the optically thick region is well established in the two-moment method. Thus, the difference of $P_{i}^{\hspace{0.5mm} j}$ between our method and the two-moment method may be small.}. On the other hand, our method does not perfectly guarantee the conservation of momentum since we keep using a original Boltzmann solver in the optically thin region. Note that if we used the prescription for the optically thick region also in the optically thin region, it would enforce the multiple conservation laws simultaneously. However, as mentioned already, that would sacrifice the accuracy and may generate different problems from the lack of precision in the moments in the optically thin region.

 Finally we make a comment on the energy feedback from neutrino to matter. Since the energy equation has nothing to do with the artificial PNS acceleration, we leave untouched the original treatment in the energy equation:
\begin{eqnarray}
&& n^{\beta} T^{\alpha}_{({\rm mat}) \beta ; \alpha} = - n^{\beta} G_{\beta}, \label{eq:enecon}
\end{eqnarray}
i.e., we compute the feedback by directly integrating the neutrino-matter interactions.

\section{Examinations} \label{sec:exam}

\begin{figure}
\vspace{15mm}
\epsscale{1.2}
\plotone{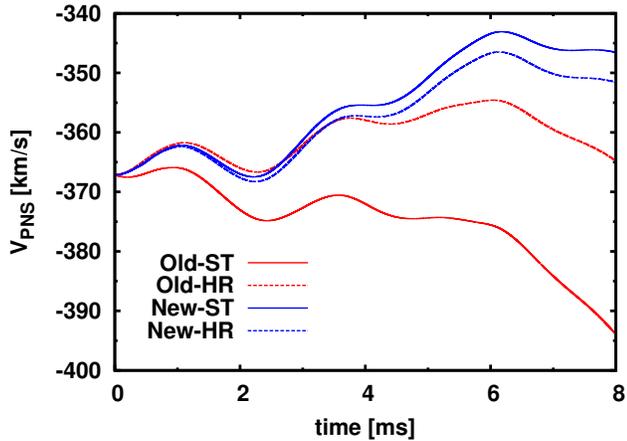}
\caption{Time evolutions of the PNS velocity ($V_{\rm PNS}$). Red and blue lines display the results of the simulations with the original and new methods, respectively. The line type indicates the difference of angular resolution in neutrino momentum space: the solid and dashed lines denote the standard and high resolutions, respectively.
\label{dcompare_VPNS200ms}}
\end{figure}

\begin{figure*}
\vspace{15mm}
\epsscale{1.2}
\plotone{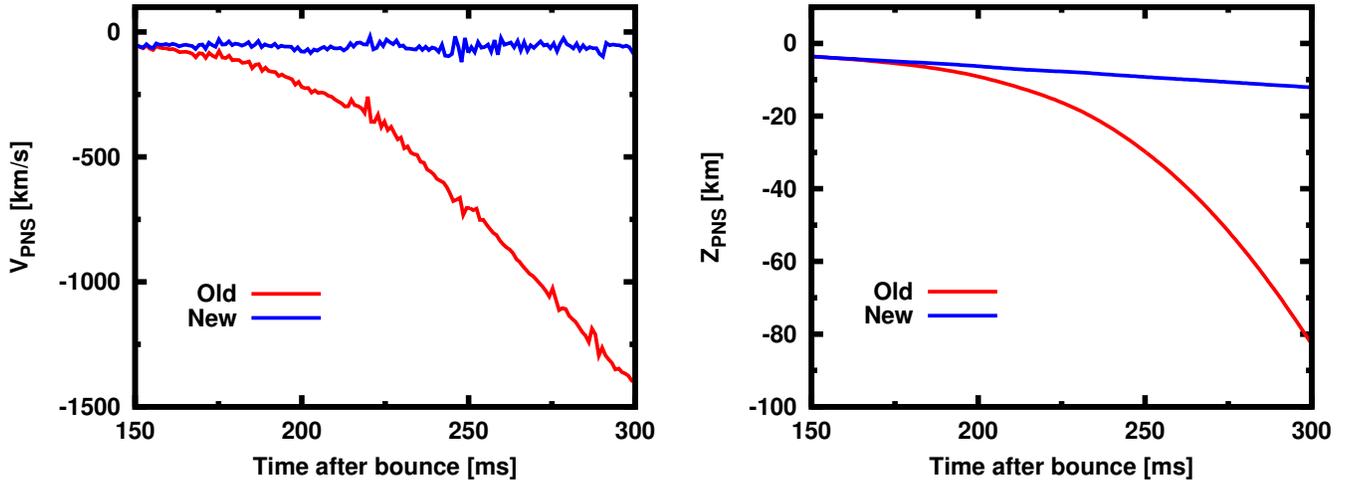}
\caption{The same as Fig.~\ref{graph_artiPNS_velo_posi} but with the result of the simulation with the new method. The red and blue lines correspond to the results of the original and new methods, respectively.
\label{graph_PNScompari_VPNS_ZPNS}}
\end{figure*}

In this section we examine the validity of our new method by running some axisymmetric CCSN simulations. As shown below, we find that the new method successfully suppresses the artificial acceleration of PNS. We also show that the new method improves the accuracy of the total linear momentum conservation in our CCSN code.

As the first examination, we run four short-term ($8$ms) simulations. We name them "Old-ST", "Old-HR", "New-ST" and "New-HR", respectively. The term of "Old" denotes the simulations under the previous treatment, while "New" is given to the models employing the new method. "ST" and "HR" represent the standard and high angular resolutions, respectively, in the former of which $10(\tilde{\theta}) \times 6(\tilde{\phi})$ and in the latter $14(\tilde{\theta}) \times 8(\tilde{\phi})$ grid points are deployed, respectively. We prepare the initial condition by taking the matter and neutrino profiles at $t=200$ms in the original simulation. In the "HR" models, we set the initial angular distribution of neutrinos by linearly interpolating the original data. In these simulations, we restore all input physics, velocity-dependent terms and the moving-mesh technique.

Fig.~\ref{dcompare_VPNS200ms} displays the time evolution of the velocity of PNS ($V_{\rm PNS}$) for all models. Different colors distinguish the previous and new treatments while the line types represent the angular resolutions. As clearly seen in this figure, our new method (blue lines) suppresses the PNS acceleration. We also find that the difference of $V_{\rm PNS}$ between New-ST and New-HR is much smaller than that in "Old" models. This indicates that the accuracy of momentum conservation is improved by our new method. Note that the result in Old-HR is close to those in "New" models. This fact implies that even the previous prescription will obtain the convergence to the true result by increasing angular resolutions. Another important finding is that New-ST may decelerate the PNS a bit excessively, since the result of New-HR comes between those of Old-HR and New-ST. This may indicate that the new treatment with the standard angular resolution would underestimate the velocity of PNS. This fact should be kept in mind in our future CCSN simulations.

Finally we carry out a long-term simulation (up to $t = 300$ms) with the new method. The simulation is started at $t=150$ms from the result of the original simulation. We choose this time because it is slightly before the start of the runaway motion of PNS in the original computation. Fig.~\ref{graph_PNScompari_VPNS_ZPNS} displays the time evolution of the position and $V_{\rm PNS}$ of PNS. We also show the same quantities in the original simulation as red lines. As can be seen in this figure, our new treatment works quite well and no numerical problems are observed in this long-term simulation.

\section{Conclusions} \label{sec:conc}
We develop a new method for the neutrino-matter coupling to improve the accuracy of the linear momentum conservation in multi-D CCSN simulations with the angular-grid-based Boltzmann solver for neutrino transport. This effort is motivated by the requirement of extremely high accuracy of CCSN simulations for reliable modelling of CCSN explosions and PNS kicks under restricted computational resources. In fact, we observed an unphysically large acceleration of PNS in one of our axisymmetric CCSN simulations and suspected that it was due to the violation of linear momentum conservation in our Boltzmann solver.

To see the problem clearly, we perform Boltzmann simulations on top of the fixed matter profile. This study reveals that the large acceleration of PNS is due to errors in the momentum feedback from neutrino to matter in the optically thick region. Note that our Boltzmann solver works basically well (see Fig.\ref{graph_steadycheck_local_v2}): indeed, it satisfies the linear momentum conservation within $\lesssim 1 \%$ for the standard angular resolution in momentum space. However, this accuracy is not sufficient to handle the PNS motion when the matter profile becomes strongly asymmetric. We find that the error of total linear momentum leads to the PNS acceleration of $\sim 10^{9} {\rm cm/s}^2$, which accounts indeed for the large acceleration of PNS observed in one of our CCSN simulations (see Fig.~\ref{graph_steady_valance}). We also find that increasing the angular resolution in momentum space alone reduces the PNS acceleration, which indicates that the inaccuracy in the momentum conservation is due to the lack of angular resolution in momentum space for neutrinos. This means that we can avoid the problem simply by increasing the angular resolution. However, it is not the best solution, since it is inhibitingly costly in terms of the numerical resource. Thus, the development of the new method for the improvement of the neutrino-matter coupling is one of necessary tasks in our CCSN project.

The basic idea in the new method is that our Boltzmann solver accurately evaluates the momentum feedback from neutrino to matter in the optically thick region with the currently available angular resolution in momentum space if the energy-momentum tensor of neutrinos is employed. This is due to the fact that the energy momentum tensor is dominated by the isotropic component of the angular distribution of neutrino. We also identify the source of errors in the previous treatment, in which the collision term is directly integrated. The problem is that it essentially utilizes the 1st moment of the angular distribution in momentum space to evaluate the momentum feedback from neutrino to matter. It should be noted that the angular-grid-based Boltzmann solvers tend to generate larger relative errors in the 1st moment than in the 0th and 2nd moments, since the former depends solely on the deviation from isotropy, which is very small in the diffusion regime; on the other hand, the latter two are mainly determined by the dominant isotropic component and can be obtained accurately in the same regime. This is the main reason why the original treatment has higher sensitivity to the angular resolution in momentum space.

Although the problem of the Boltzmann neutrino transport in the diffusion regime has been already recognized and possible solutions were proposed for some numerical schemes (see e.g., \citet{2016ApJ...817...72P}), they took full advantage of the diffusion approximation, which is not compatible with our Boltzmann solver. Moreover, we need to ensure that the solution should be consistent with the general relativistic formulation of our code. This is not a trivial issue and needs the new method, which satisfies all these requirements.

In the new method, the momentum feedback from neutrino to matter is computed based on the temporal and spatial derivatives of energy-momentum tensor (see Eq.~(\ref{eq:momentumexpression})). Although the new treatment can be applied in all regimes in principle, we adopt it only in the optically thick region. In the optically thin region, we continue to use the original treatment, in which we compute the momentum exchange between neutrino and matter by integrating the collision terms for individual neutrino-matter interactions and then give their sum to the Euler equation. In fact, the original method is more suitable in the optically thin region, since the momentum exchange decreases just as expected as the neutrino-matter interaction becomes inefficient, while it is not necessarily guaranteed in the new treatment.

We find that the new method works quite well. It successfully suppresses the unphysically large PNS acceleration even with the standard angular resolution in momentum space (see Fig.~(\ref{graph_PNScompari_VPNS_ZPNS})). The angular-resolution study also shows that the difference between the standard- and high-resolution simulations is smaller in the new method than in the original treatment (see Fig.~(\ref{dcompare_VPNS200ms})).

Importantly, we still have a coherent PNS motion in the same simulation but with the new treatment (see Fig.~\ref{graph_PNScompari_VPNS_ZPNS}). We believe that this recoil of PNS is real, i.e., this is the first successful direct treatment of PNS kick in multi-D radiation-hydrodynamic simulations with the full Boltzmann neutrino transport. The impact on CCSN dynamics will be discussed in a separate paper. Last but not least, the new method will be useful to improve the accuracy of the momentum exchange between radiation and matter for other radiation-hydrodynamic codes that adopt multi-angle treatments such as those by \citet{2004ApJ...609..277L,2008ApJ...685.1069O} for neutrinos and by \citet{2014ApJS..213....7J,2014ApJ...796..106J,2015ApJS..217....9R,2018MNRAS.tmp.2941H} for photons. The versatility is one of the important merits in the new method.

\acknowledgments 
 We acknowledge Adam Burrows for fruitful discussion. The numerical computations were performed on the supercomputers at K, at AICS, FX10 at Information Technology Center of Nagoya University. Large-scale storage of numerical data is supported by JLDG constructed over SINET4 of NII. H.N. was supported by Princeton University through DOE SciDAC4 Grant DE-SC0018297 (subaward 00009650). This work was also supported by Grant-in-Aid for the Scientific Research from the Ministry of Education, Culture, Sports, Science and Technology (MEXT), Japan (15K05093, 25870099, 26104006, 16H03986, 17H06357, 17H06365), HPCI Strategic Program of Japanese MEXT and K computer at the RIKEN (Project ID: hpci 160071, 160211, 170230, 170031, 170304, hp180179, hp180111, hp180239).
\bibliography{bibfile}

\end{document}